# Reverse degradation of nickel graphene junction by hydrogen annealing

Zhenjun Zhang, Fan Yang, Pratik Agnihotri, Ji Ung Lee, J. R. Lloyd[1] [*]

College of Nanoscience and Engineering, SUNY Polytechnic Institute, Albany, NY USA 12203

Metal contacts are fundamental building components for graphene based electronic devices and their properties are greatly influenced by interface quality during device fabrication, leading to resistance variation. Here we show that nickel graphene junction degrades after air exposure, due to interfacial oxidation, thus creating a tunneling barrier. Most importantly, we demonstrate that hydrogen annealing at moderate temperature (300 $^0$C) is an effective technique to reverse the degradation.

Graphene is a promising material in the application of high speed field effect transistors (FETs) and many efforts have been focused on the study of the intrinsic transport property of graphene.[1,2,3,4] While the maximum resistivity at charge neutral point has been predicted to have a universal value (off state), the on state conductivity of graphene FETs are dictated by metal graphene contact.[5] Contact resistance is parasitic and has become a limiting factor for aggressively scaling CMOS technology.[6] As a candidate for post CMOS material, it is important to fully understand the behavior of different metal candidates in contacting with graphene.

To date, a handful of metals have been studied and the values of contact resistivity ($\rho_c$) showed large variation from a few hundred ohms per μm to tens of kilo-ohms per μm, even for the same metal specie.[7,8,9,10,11,12,13,14,15,16,17] It has been argued that these variation can be attributed to the quality of graphene (exfoliated, CVD grown, epitaxial grown), the type of the contact (top or edge conduction), pre and post fabrication treatment as well as the underlying substrate.[18] Among all these metals, Pd metal contacts have been consistently reported to have a resistivity on the order of 100 ohms μm.[8,19] Ni and Ti

---

[1] Electronic address: jlloyd@sunypoly.edu; Present Address: SUNY Polytechnic Institute, Albany, New York 12203, USA

contacts have also been reported to reach a lower contact resistivity but the values are strongly process specific.[20,13] The scarcity of Pd metal limits its application in large scale manufacturing, so it becomes critical to understand the origin of the resistance variation at nickel and titanium contacts.

Theoretical studies have suggested two important factors impacting the contact resistivity: carrier injection from the metal graphene vertical junction and the potential profile near the longitudinal edge.[8] While there is a fundamental maximum resistivity at the graphene pn junction, the origin of the large variation on the nickel and titanium contacts are presumably from the vertical carrier injection.[21] It is well known that the transfer process of CVD grown graphene and lithography processes have been shown to introduce residues, creating an interface layer.[22] The existence of interfacial layer can 1. Reduce the doping effect from the metal to graphene. 2. Increase the carrier tunneling distance at the junction.[23]. Assuming an insulating behavior, the carrier tunneling probability is exponentially dependent on the thickness of the interfacial layer.[21] Therefore, the existence of interfacial layer can be a key factor for inconsistent resistivity values.

While many efforts have been made to optimize the process condition to reduce interfacial contamination,[24,14,25,26,27,28] little attention has been paid to interface degradation post fabrication. Xu et al. observed an intentional degradation on aluminum graphene contact by metal evaporation under oxygen environment.[29] Nouchi et al. observed a depinning effect near the charge neutral point under a two terminal measurement using nickel and cobalt as electrodes, speculating oxidation at the metal graphene interface.[30, 31] Here we studied a nickel graphene junction and characterize the interface by Auger and XPS spectroscopy, identifying the origin of the deterioration. More importantly, we demonstrate hydrogen annealing is an effective technique to reverse the performance of nickel graphene junction. A similar degradation was also observed on titanium graphene junction but the change of palladium graphene junction is limited at the channel, consistent with the most recent report.[18]

In the study, CVD graphene is used and the transfer process is reported elsewhere.[32] To minimize the impact of organic residue contamination during the transfer process, a low temperature baking at 150 $^0$C at high vacuum (~$10^{-7}$ torr) was used post graphene transfer. Back-gated graphene field effect transistor devices were fabricated using a conventional photolithography process. First, graphene was patterned using a positive photoresist, followed by etching of unwanted graphene regions by oxygen plasma. Second, the electrodes were patterned on graphene using image reversal photoresist, followed by metal evaporation and lift-off process. All electrical data were measured using Keithley S4200 in a Lakeshore vacuum-cryostat probe station at room temperature in vacuum atmosphere of $10^{-5}$ Torr. Both four and two terminal probes were used to measure the change of resistance post fabrication and after two weeks of ambient exposure. Then, devices were carefully loaded into a CVD chamber with Ar: $H_2$ (4:1) for 1 hour at ambient pressure, gradually increase the temperature to 300$^0$C and maintain for 1 h with cooling down by convection.

To understand the impact of hydrogen annealing on the both graphene channel and the metal junction, both four point kelvin probe structures and standard transmission line structures were fabricated in the same die. Figure 1a shows a representative $I_d$-$V_g$ curve for the graphene channel at different stage after fabrication. Two key parameters, the mobility and minimum conductance point were extracted using a method outlined elsewhere.[18] Figure 1(b) summarized the mobility change at different stage of tests. The as-fabricated devices have an average mobility of 4700 cm$^2$V$^{-1}$s$^{-1}$. After air exposure, the devices saw an increase in average mobility up to 6000 cm$^2$v$^{-1}$s$^{-1}$, suggesting decreasing of charge induced impurities.[33] This could attribute to desorption of volatile molecules that is trapped at graphene surface or substrate graphene interface. After hydrogen annealing, the mobility degrades to 3900 cm$^2$v$^{-1}$s$^{-1}$. The impact of hydrogen annealing in graphene channel is not fully understood. While some literature reports a reversible degradation due to weak hydrogen graphene interaction,[34] others claim improvement of channel mobility.[35] These suggest that the role of hydrogen annealing on a fully functional FET device with substrate can be multifold. In our case, the degradation can be attributed to temperature induced coupling between $SiO_2$ substrate and graphene at elevated temperature.[36] Figure 1(c) shows the shift of charge neutral point on the same group of devices. In summary, all the fabricated devices showed moderate Dirac point shift. The as-fabricated devices showed an average voltage shift of 2.2 V. The air exposed and $H_2$ annealed devices showed an average voltage shift of 1.6 and 0.6 V, suggesting a decrease in charge inhomogeneity on the graphene surface. It is calculated that for a 300 nm $SiO_2$ substrate, 1 V of back-gated voltage induce about 0.03 eV Fermi level shift in graphene. Therefore, the observed voltage shift is essentially rather small in all three cases.

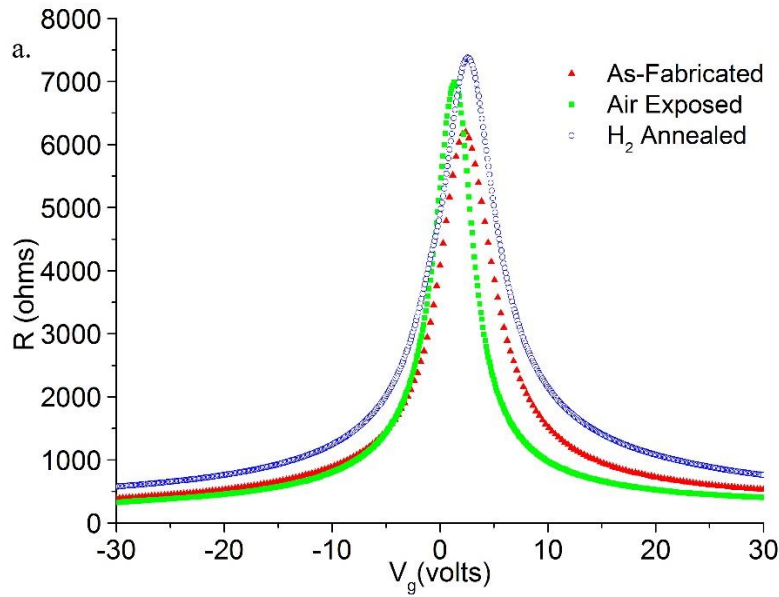

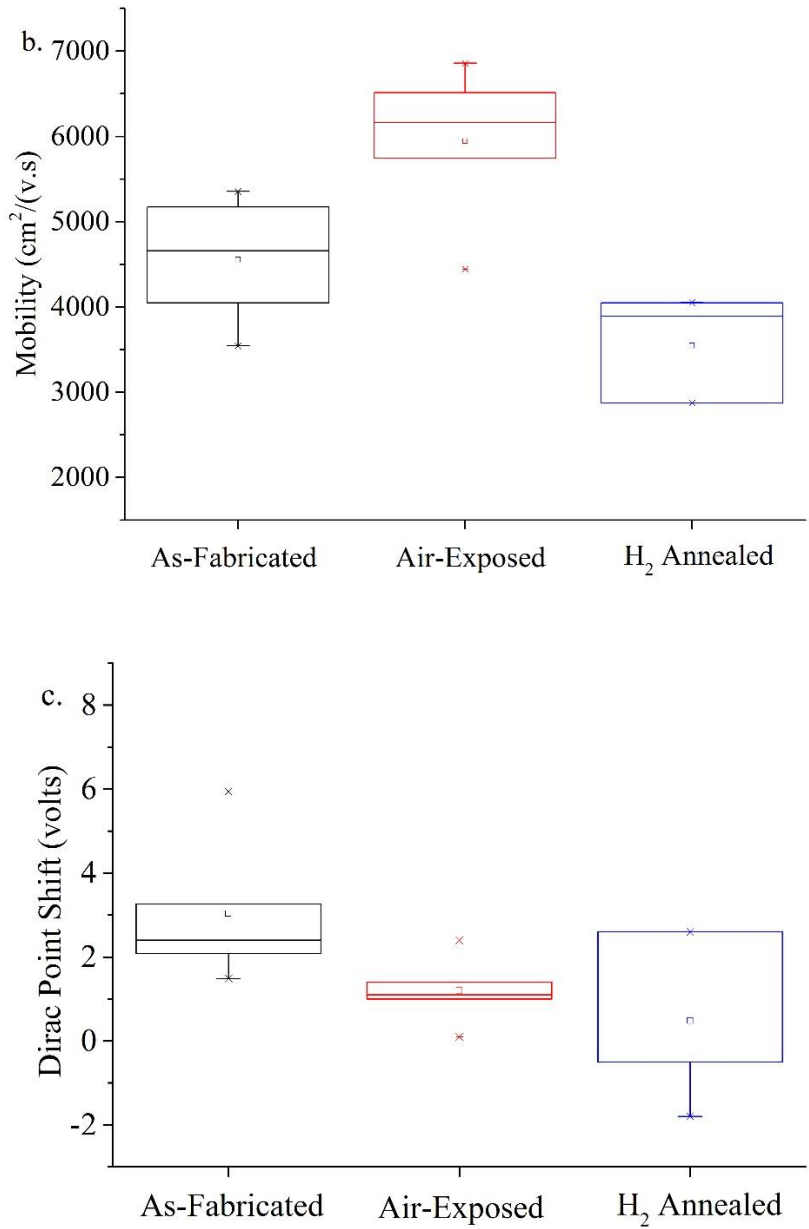

Figure 1. (Color online) a) Representative four point measurement I-V characteristics. b) Mobility measurement. c) Charge neutral point shift measurement.

To further characterize the impact of $H_2$ annealing on the metal graphene junction, devices with transmission line structures were tested. Figure 2a shows typical I-V behavior at different stage of testing. In these testing structures, the measurement resistance value can be simply described as $R_{total}=2R_{ex}+2R_c+R_{ch}$, where $R_c$ is the contact resistance, $R_{ch}$ is the channel resistance and $R_{ex}$(~20 ohms) accounts for the resistance contribution from the probe to the metal, which can be negligible in our case. As illustrated in Figure 1, the impact of hydrogen annealing is limited in the channel, thus the total resistance change in a two terminal measurement is dominant by contact resistance. At channel carrier density (n) of $2.2 \times 10^{12}$ cm$^{-2}$, the plotted initial residue resistance is 816 ohms.μm. After air exposure for 2 weeks, more than 4 fold degradation was observed with measured resistance of 3600 ohms.μm. After hydrogen annealing for 1h, the resistance is reduced by more than 14 fold, to 250 ohms.μm, indicating a significant improvement at metal graphene contact. Additional annealing time (up to 4h) does not further improve the contacts, indicating no significant change at the interface.

For an appropriate extrapolation of contact resistance, transmission line measurement (TLM) was used for the devices treated with hydrogen annealing. Figure 2b shows a set of typical transmission line measurement data. The contact resistance is extracted from different channel at several representative voltages. Consistent with previous report on nickel graphene contact, the contact resistance values are dependent on the back-gate voltage, indicating a gate modulation happens at the junction.[5] It is worth mention that near the charge neutral point, the total resistance are dominant by the channel, resulting in deviation in contact resistance extrapolation.[19] Meanwhile, it is rather challenging to extract contact resistance from the degraded devices, suggesting non-uniform interfacial oxidation after air exposure. Instead, kelvin probe measurements were used to estimate the contact resistivity. Table I shows the summary of contact resistivity obtained for three different metal stacks (Ti/Au, Pd/Au, Ni) at n=$2.2 \times 10^{12}$ cm$^{-2}$. In contrast to Ti/Au and Ni devices, Pd showed negligible degradation after air exposure. Echoing with recent report, titanium can be oxidized at the graphene junction while the palladium contacts are inert to oxidation.[18] However, it is worth noting that both Pd and Ti/Au contacts were vulnerable to hydrogen annealing in this set of experiment, suggesting care needs to be taken when hydrogen is involved in fabricating multilayer device structures.

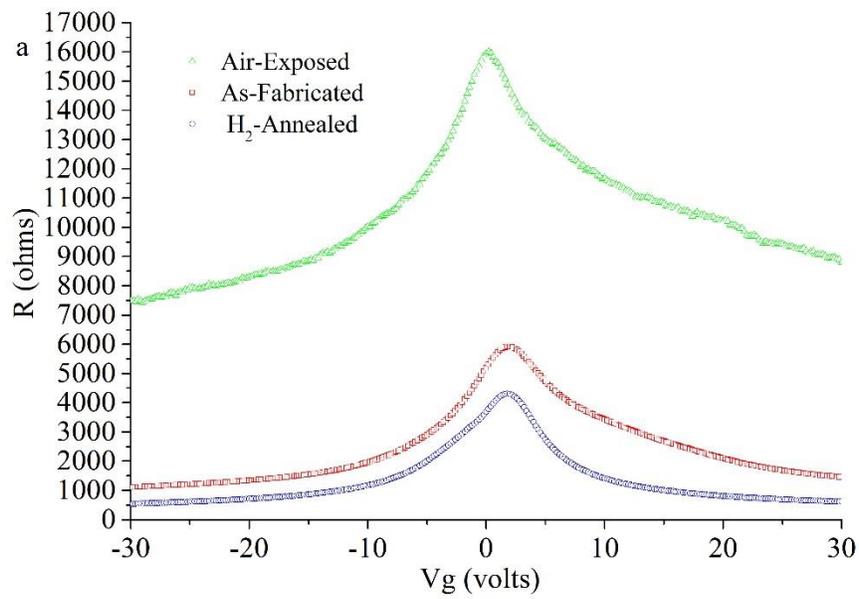

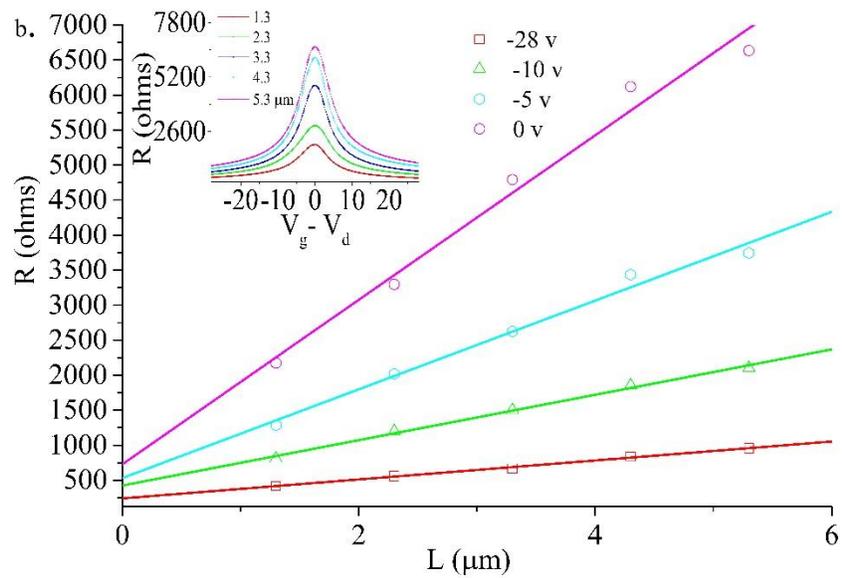

Figure 2. (Color online) (a) Representative 2 terminal measurement results at different stage of the test, (b) Contact resistance extraction after hydrogen annealing at different gate bias(inset: I-V curve for different channel length as a function of gate bias).

TABLE I. Contact resistance for different metal graphene contacts ($n=2.2 \times 10^{12} cm^{-2}$).

| Unit: ohms. μm | As-Fabricated | Air-Exposed | $H_2$-Annealed |
|:---:|:---:|:---:|:---:|
| 5 nm Ti/45 nm Au | ~469 | ~3000 | X |
| 50 nm Ni | ~854 | ~6500 | 232±44 |
| 40 nm Pd/10 nm Au | 200±52 | 169±43 | X |

To further characterize the nickel graphene interface, a depth profile of Auger Spectroscopy was performed to investigate the chemical changes before and after hydrogen annealing. Carbon associated with graphene (in blue) was identified at Ni-$SiO_2$ interface. At the interface, there is a sharp decrease in nickel concentration and abrupt increase in silicon and oxygen concentration. Comparing the two profiles at interface in parallel (Figure 3a), a small amount of oxygen was accumulated near graphene nickel interface, suggesting nickel oxide formation. The nickel to oxygen ratio is roughly 9:1 at the peak position, indicating a possible non uniform oxidation. To further confirm the oxygen is from nickel oxide, XPS depth profile analysis was used to review the chemical bonding information. The data presented in Figure 4c is a fraction of the O 1s profile at the interface of nickel and $SiO_2$. A peak at 529.9 eV is assigned to NiO, and a broadened peak at 332.1 eV were assigned to a mixture signal from $SiO_2$ and defective oxygen peak in $NiO_x$. After hydrogen annealing, the extra amount of oxygen concentration in the profile was effectively removed, resulting pure $SiO_2$ peaks.

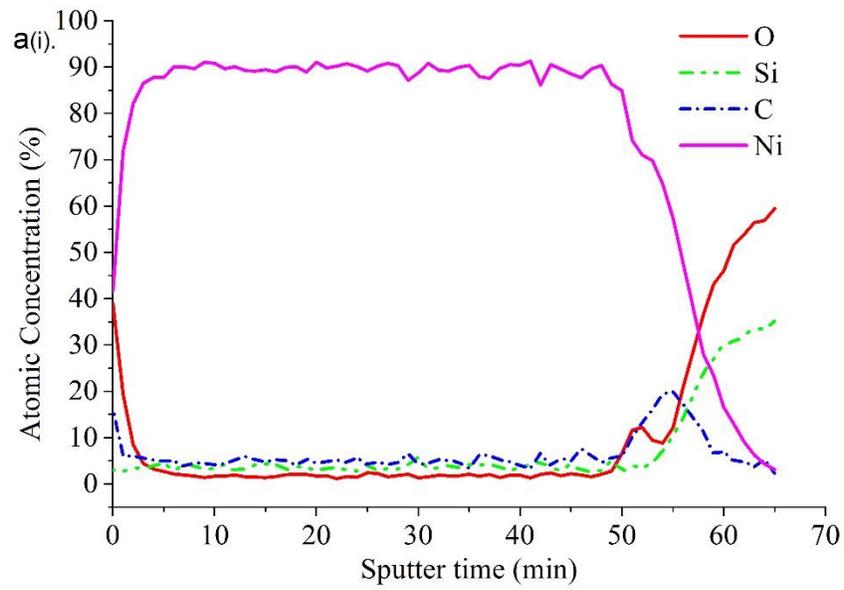

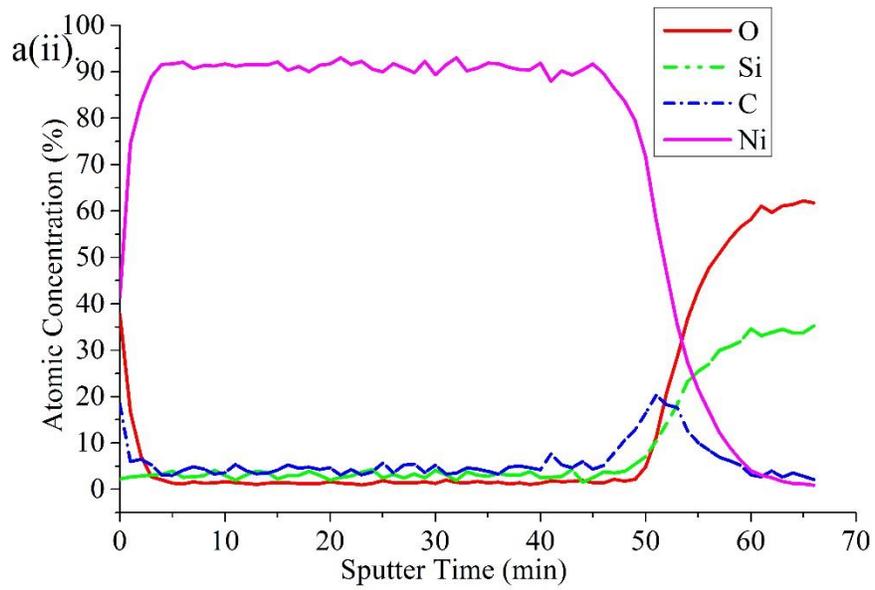

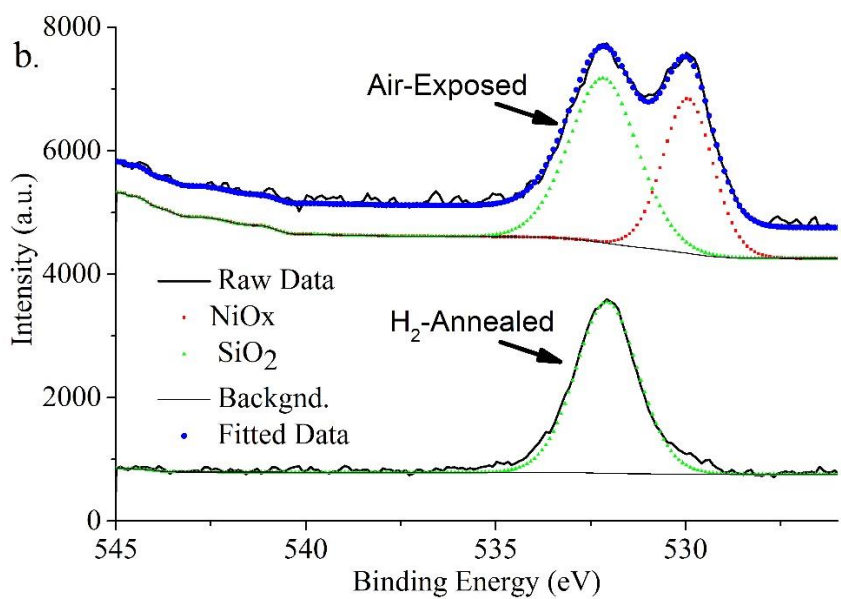

Figure 3. (Color online) (a) Auger depth profile of Ni/Gr/SiO2/Si (i) Air-Exposed, (ii) $H_2$-Annealed (b) O1s X-ray photoelectron peaks near Ni-Gr-$SiO_2$ interface.

The correlation between interface oxidation, hydrogen reduction and the electrical properties at nickel graphene junction are illustrated in Figure 4. After air exposure, the nickel graphene junction was separated by an oxide layer, evidenced by Auger and XPS measurement. Other than reduced charge transfer at the interface, the oxide layer served as a tunneling barrier, increasing contact resistivity. Using hydrogen annealing, nickel oxide is reduced to metallic nickel and the barrier layer is removed with enhanced dipole interaction, resulting in significant improvement on the contact.

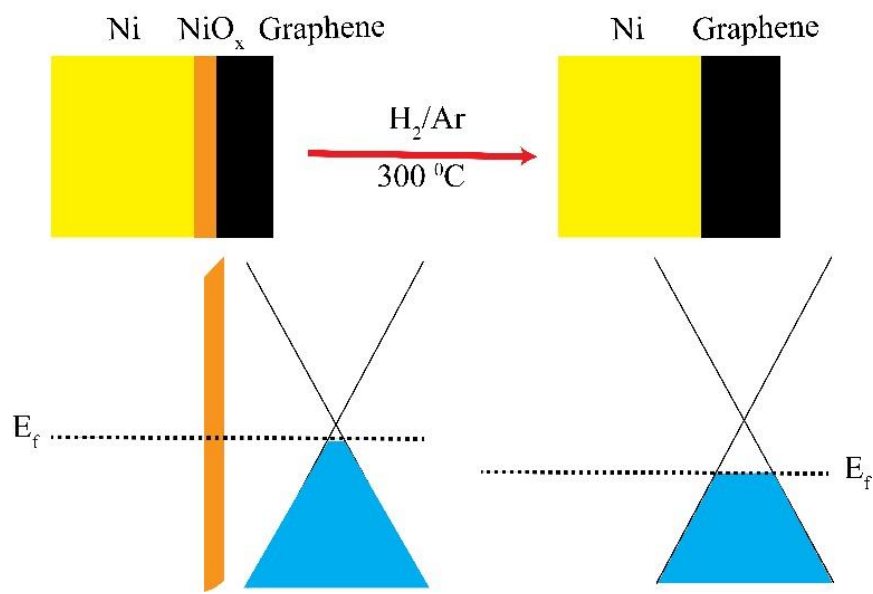

Figure 4. Correlation between interface redox chemistry and band diagram evolution.

In summary, oxyphilic metals (Ni, Ti) are vulnerable to oxidation at graphene metal interface and result in formation of non-uniform tunneling barrier, thus increasing the contact resistance after air exposure. We have shown that hydrogen annealing can be an effectively method to reverse degradation at the graphene nickel interface, improving contact properties.

Z. Zhang is grateful for research assistantship support from College of Nanoscience and Engineering at SUNY Polytechnic Institute.

[1] K. S. Novoselov, A. K. Geim, SV Morozov, D. Jiang, Y. Zhang, SV Dubonos, IV Grigorieva, and AA Firsov, Science, **306**, 666 (2004).
[2] I. Meric, M. Y Han, A. F. Young, B. Ozyilmaz, P. Kim, and K. L. Shepard, Nat. Nanotechnol., **3**, 654 (2008).

3 F. Xia, D. B. Farmer, Y. Lin, and P. Avouris, Nano Lett. **10**, 715 (2010).
4 S. Kim, J. Nah, I. Jo, D. Shahrjerdi, L. Colombo, Z. Yao, E. Tutuc, and S. K. Banerjee, arXiv preprint arXiv:0901.2901 (2009).
5 K. Nagashio, T. Nishimura, K. Kita, and A. Toriumi, Appl. Phys. Lett. **97**, 143514 (2010).
6 Q. Cao, S. Han, J. Tersoff, A. D Franklin, Y. Zhu, Z. Zhang, G. S. Tulevski, J. Tang, and W. Haensch, Science **350**, 68 (2015).
7 B. Huang, M. Zhang, Y. Wang, and J. Woo, Appl. Phys. Lett. **99**, 032107 (2011).
8 F. Xia, V. Perebeinos, Y. M. Lin, Y. Wu, and P. Avouris, Nat. Nanotechnol., **6**, 179 (2011).
9 E. Watanabe, A. Conwill, D. Tsuya, and Y. Koide, Diamond Relat. Mater., **24**, 171 (2012).
10 A. Venugopal, L. Colombo, and E. M. Vogel, Appl. Phys. Lett. **96**, 013512 (2010).
11 S. Russo, M. F. Craciun, M. Yamamoto, A. F. Morpurgo, and S. Tarucha, Physica E, **42**, 677 (2010).
12 J. S. Moon, M. Antcliffe, H. C. Seo, D. Curtis, S. Lin, A. Schmitz, I. Milosavljevic, A. A. Kiselev, R. S. Ross, D. K. Gaskill, P. M. Campbell, R. C. Fitch, K.-M. Lee, and P. Asbeck, Appl. Phys. Lett. **100**, 203512 (2012).
13 J. A. Robinson, M. LaBella, M. Zhu, M. Hollander, R. Kasarda, Z. Hughes, K. Trumbull, R. Cavalero, and D. Snyder, Appl. Phys. Lett. **98**, 053103 (2011).
14 L. Wang, I. Meric, P. Y. Huang, Q. Gao, Y. Gao, H. Tran, T. Taniguchi, K. Watanabe, L. M. Campos, D. A. Muller, J. Guo, P. Kim, J. Hone, K. L. Shepard, and C. R. Dean, Science **342**, 614 (2013).
15 A. Hsu, H. Wang, K. K. Kang, J. Kong, and T. Palacios, EDL, IEEE, **32**, 1008 (2011).
16 J. Lee, Y. Kim, H. Shin, C. Lee, D. Lee, C. Moon, J. Lim, and S. C. Jun, Appl. Phys. Lett. **103**, 103104 (2013).
17 M. Politou, I. Asselberghs, I. Radu, T. Conard, O. Richard, C. S. Lee, K. Martens, S. Sayan, C. Huyghebaert, and Z. Tokei, Appl. Phys. Lett. **107**, 153104 (2015).
18 H. Zhong, Z. Zhang, B. Chen, H. Xu, D. Yu, L. Huang, and L. Peng, Nano Res. **8**, 1669 (2015).
19 W. S. Leong, H. Gong, and J. TL Thong, ACS nano **8**, 994 (2013).
20 J.S. Moon, M. Antcliffe, H.C. Seo, D. Curtis, S. Lin, A. Schmitz, I. Milosavljevic, A.A. Kiselev, R.S. Ross, and D.K. Gaskill, Appl. Phys. Lett. **100**, 203512 (2012).
21 D. Berdebes, T. Low, Y. Sui, J. Appenzeller, and M. S. Lundstrom, IEEE Trans. Electron Devices, **58**, 3925 (2011).
22 Y.C. Lin, C. C. Lu, C.H. Yeh, C. Jin, K. Suenaga, and P. W. Chiu, Nano Lett. **12**, 414 (2011).
23 R. Ifuku, K. Nagashio, T. Nishimura, and A. Toriumi, Appl. Phys. Lett. **103**, 033514 (2013).
24 O. B. and C. Kocabas, Appl. Phys. Lett. **101**, 243105 (2012).
25 W. S. Leong, H. Gong, and J. T. L. Thong, ACS Nano **8**, 994 (2014).
26 C. W. Chen, F. Ren, G. C. Chi, S.C. Hung, Y. P. Huang, J. Kim, I. I. Kravchenko, and S. J. Pearton, J. Vac. Sci. Technol., B **30**, 060604 (2012).
27 W. Li, Y. Liang, D. Yu, L. Peng, K. P. Pernstich, T. Shen, AR. H.Walker, G. Cheng, C. A. Hacker, and C. A. Richter, Appl. Phys. Lett. **102**, 183110 (2013).
28 M. S. Choi, S. H. Lee, and W. J. Yoo, J. Appl. Phys. **110**, 073305 (2011).
29 X. Du, I. Skachko, and E. Y. Andrei, Int. J. Mod. Phys. B, **22**, 4579 (2008).
30 R. Nouchi, M. Shiraishi, and Y. Suzuki, arXiv preprint arXiv:0810.3057 (2008).
31 R. Nouchi and K. Tanigaki, Appl. Phys. Lett. **96**, 253503 (2010).
32 Z. Hu, D. P. Sinha, J. U. Lee, and M. Liehr, J. Appl. Phys., **115**, 194507 (2014).
33 J.H. Chen, C. Jang, S. Adam, M.S. Fuhrer, E.D. Williams, and M. Ishigami, Nat. Phys., **4**, 377 (2008).
34 D.C. Elias, R.R. Nair, T. Mohiuddin, SV. Morozov, P. Blake, M.P. Halsall, A.C. Ferrari, D.W. Boukhvalov, M.I. Katsnelson, and A.K. Geim, Science, **323**, 610 (2009).

35   W. S. Leong, C. T. Nai, and J. T. Thong,  Nano Lett., **14**, 3840 (2014).
36   J.H. Chen, C. Jang, S. Xiao, M. Ishigami, and M. S. Fuhrer,  Nat. Nanotechnol., **3**, 206 (2008).